\documentstyle[12pt]{article}

\catcode`\@=11
\def\numberbysection{\@addtoreset{equation}{section}
	\def\theequation{\thesection.\arabic{equation}}}
\def\beq{\begin{equation}}
\def\eeq{\end{equation}}
\def\barr{\begin{eqnarray}}
\def\earr{\end{eqnarray}}
\def\winf{W_{1+\infty}\ }
\begin{document}
\begin{titlepage}
\begin{center}
\hfill DFTT 18/95 \\
\hfill hep-th/9503229 \\
\vskip .3 in
{\large \bf The $\winf$ effective theory of the Calogero-Sutherland \\
 model and Luttinger systems }
\vskip 0.2in
Raffaele CARACCIOLO \\
{\em I.N.F.N. and Dipartimento di Fisica Teorica,
          Universit\`a di Torino,\\
	  Via P. Giuria 1, I-10125 Torino, Italy}
\\
\vskip 0.1in
Alberto LERDA \\
{\em Dipartimento di Fisica Teorica,
     Universit\`a di Salerno,\\
     I-84100 Salerno, Italy,
     and I.N.F.N., Sezione di Napoli}
\\
\vskip 0.1in
Guillermo~R.~ZEMBA \\
{\em I.N.F.N. and Dipartimento di Fisica Teorica,
          Universit\`a di Torino,\\
	  Via P. Giuria 1, I-10125 Torino, Italy}
\end{center}
\vskip .1 in
\begin{abstract}
\noindent
We construct the effective field theory of the Calogero-Sutherland
model in the thermodynamic limit of large number of particles $N$.
It is given by a $\winf$ conformal field theory (with
central charge $c=1$) that describes {\it exactly} the spatial density
fluctuations arising from the low-energy excitations about the
Fermi surface.
Our approach does not rely on the integrable character of the
model, and indicates how to extend previous results to any order
in powers of $1/N$.
Moreover, the same effective theory can also be used to describe
an entire universality class of $(1+1)$-dimensional fermionic
systems beyond the Calogero-Sutherland model, that we identify
with the class of {\it chiral Luttinger systems}.
We also explain how a systematic bosonization procedure can be
performed using the $\winf$ generators, and propose this algebraic
approach to {\it classify} low-dimensional non-relativistic fermionic
systems, given that all representations of $\winf$ are known.
This approach has the appeal of being mathematically complete and
physically intuitive, encoding the picture suggested by Luttinger's
theorem.
\end{abstract}

\vfill
\hfill March 1995
\end{titlepage}
\pagenumbering{arabic}
%

The Calogero-Sutherland model \cite{cal}\cite{sut}
has recently received considerable
attention as a testing model for new ideas in low-dimensional
systems significant for condensed matter and theoretical physics.
Among the most exciting new ideas we find Haldane's fractional
statistics in $(1+1)$-dimensional systems \cite{hald2}. The
model has been also used in connection with algebraic approaches
based on $\winf$ algebras \cite{csw}, applications to
two-dimensional $QCD$ \cite{qcd}, integrable spin-chains with long-range
interactions \cite{chain}, anyons in a magnetic field \cite{anyon},
quantum wires \cite{quaw}, the quantum Hall effect \cite{csqhe},
string theory and matrix models \cite{jevic} and possibly others.

In this letter we construct the low-energy {\it effective field theory}
\cite{polch} of the model in the
thermodynamic limit, in terms of the generators of the $\winf$ dynamical
symmetry that describe the lowest energy (``gapless'') spatial density
fluctuations of the many-body states of the system. This construction is
a straightforward application of a method successfully developed for
the quantum Hall effect \cite{cdtz}\cite{ctz4}. Indeed,
writing the model in terms of the $\winf$ generators allows
us to include the non-perturbative effects of the interaction
by simply changing the representation of the symmetry algebra,
a result implied by Luttinger's theorem \cite{hald1}.
The resulting effective conformal field theory \cite{bpz}
incorporates, therefore, the idea of the {\it bosonization} of the
lowest energy {\it fluctuations of the Fermi surface} as the relevant
semiclassical degrees of freedom \cite{boson}. Our results expand
upon previous findings \cite{kaya} and, moreover, show that
the same theory describes general Luttinger systems as well.

Previous investigations in the Calogero-Sutherland model using
conformal and $\winf$ methods \cite{csw}
have been mostly concerned with the integrability aspects of
the model encoded in the symmetry algebra.
Here, however, we study the effective theory
of the dynamics of the Fermi surface, giving up integrability
in exchange for a formulation that applies to a
wider universality class of theories beyond the Calogero-Sutherland
model: the Luttinger systems.
With respect to the Fermi surface dynamics, the former approaches
should be considered as ``bulk'' (deep interior of the Fermi sea)
formulations, as opposed to the ``boundary'' description
encoded in the latter. Indeed, from the
experience drawed from the quantum Hall effect \cite{prange},
we expect that the ``boundary'' description will produce the most
simple and universal approach to the low-energy dynamics for these
systems \cite{ctz4}.

Consider first the non-relativistic $N$-body problem of
$(1+1)$-dimensional spinless interacting fermions on a circle of length
$L$, with hamiltonian \cite{sut} (in units where $\hbar=2m=1$,
with $m$ being the mass of the particles)
\beq
h_{CS}=\sum_{j=1}^N\ \left( \frac{1}{i} \frac{\partial}{\partial x_j}\
\right)^2\ +\ g\ \frac{\pi^2}{L^2}\ \sum_{i<j}\ \frac{1}{\sin^2
(\pi(x_i-x_j)/L) }\ .
\label{ham}
\eeq
Here $x_i$ ($i=1,\dots,N$) is the coordinate of the $i$-th particle,
and $g \equiv 2 \lambda (\lambda -1)$ the coupling constant.
We shall construct the effective field theory of (\ref{ham}) following
closely the method developed for the case of quantum Hall effect
\cite{cdtz}\cite{ctz4}, {\it i.e.}, we reformulate the problem in variables
that describe directly the dynamics of the Fermi surface when $N$ is large.
In the free case ($\lambda=1$), the single-particle wave functions
are given by plane waves:
\beq
\phi_k (x)\ = \frac{1}{\sqrt{L}}\ \exp\ \left( i \frac{2\pi}{L} k x
\right)\ ,
\label{wfcir}
\eeq
where $k$ is an integer (half-integer) for periodic (anti-periodic)
boundary conditions.
We then define a second-quantized non-relativistic
fermion field as
\beq
\Psi (x,t)\ \equiv\ \sum_{k=-\infty}^{\infty}\ a_k\  \phi_k (x,t)\  ,
\qquad \{\ a_k , a^{\dag}_l\ \}\ =\ \delta_{k,l}\ .
\label{field}
\eeq
Here $\phi_k (x,t)=\phi_k (x)\exp\left(-i {\epsilon}_k t \right ) $,
${\epsilon}_k = (2\pi/L)^2 k^2$
and $a_k$, $a^{\dag}_l$ are Fock space operators.
The ground state of the system is
\beq
|\ \Omega\ , N\ \rangle\ =\ a^{\dag}_{-M} a^{\dag}_{-M+1}\dots
a^{\dag}_{M-1}a^{\dag}_{M} |\ 0\ \rangle\ ,
\label{vac}
\eeq
where $|\ 0\ \rangle$ is the Fock vacuum and $M \equiv (N-1)/2$.
Note that $k$ is an integer (half-integer) if $N$ is odd (even).
The ground state density $\rho(x,t)$ is given by
\beq
\rho(x,t)\ \equiv\
\langle\ \Omega\ , N\ |\Psi^{\dag} (x,t) \Psi (x,t)
|\ \Omega\ , N\ \rangle\ =\ \sum_{k=-M}^{M}
|\phi_k(x,t) |^2\ =\ \frac{N}{L}\ .
\label{gsd}
\eeq
Therefore $\rho(x,t)=\rho_0=N/L$ is uniform and stationary.
We shall consider the thermodynamic limit of large $N$ and $L$,
with $\rho_0$ finite. The Fermi momentum is
$p_F=\pi(N-1)/L$ and the Fermi surface consists of two Fermi points
located at $\pm p_F$. Note that in this limit the ground-state
(\ref{vac}) becomes a relativistic Dirac sea for each one of them.
Next we define ``shifted'' Fock operators around each Fermi point
\cite{cdtz}\cite{hald1}\cite{boson} $b^{(+)}_r\ \equiv\ a_{M+r}\ $,
$b^{(-)}_r\ \equiv\ a_{-M-r}\ $, where $ |r| \ll N$ describes small
fluctuations, {\it i.e.}, the lowest energy excitations
of the system.
In the thermodynamic limit, after rescaling $x=q L/2\pi$ and
$t= \tau L/4\pi$, we find
\barr
\Psi (x,t)\ & = &\ {\frac{1}{\sqrt{L}}}\ \left[\ {\rm e}^{i\gamma(+)}\
F^{(+)} (q-p_F \tau )\ + \ {\rm e}^{-i\gamma(-)}\
F^{(-)} (q+p_F \tau )\ \right]\ , \nonumber\\
F^{(\pm)} (\theta)\ & = &\ \sum_{r=-\infty}^{\infty} b^{(\pm)}_r\
{\rm e}^{\pm ir\theta}\ ,\qquad \gamma (\pm)\ \equiv\
p_F L(q \mp p_F\tau/2)/2\pi\ ,
\label{efop}
\earr
where $F^{(\pm)}(\theta)$ are Weyl fermion
fields \cite{bpz}, which describe the relevant degrees of freedom
of the system in the vicinity of the Fermi surface \cite{polch}.
Note that in the thermodynamic limit
 the original single field (\ref{field}) decomposes
into two independent\footnote{
Apart from the overall constraint of total charge (or
particle number) conservation.}
fields of opposed chirality (\ref{efop})
(under the hypothesis of small
fluctuations around the Fermi points).
Each field represents a charged, chiral, relativistic fermion that
corresponds to a conformal field theory with central charge $c=1$
\cite{kaya}. Note that $F^{(\pm)} (\theta + 2\pi) =
\exp(\pm i 2\pi \mu) F^{(\pm)} (\theta)$, with $\mu=0$ ($\mu=1/2$)
for $N$ odd (even).
As a consequence of the above decomposition, the Hilbert space
of the effective theory becomes the tensor product of two
independent Hilbert spaces $(\pm)$.

In terms of these fields, we can immediately construct two
commuting sets of $\winf$ generators \cite{ctz3}\cite{ctz4}.
For simplicity, we shall concentrate on one of them only,
and omit the affix $(\pm)$ in the rest of the discussion
\begin{eqnarray}
V_n^j\ &=&\ \int_0^{2\pi} d\theta\
F^{\dag}(\theta)\ \ddagger {\rm e}^{-in\theta}\left(i\partial_{\theta}
\right)^j \ddagger\ F(\theta)\ , \nonumber\\
\ &=&\ \sum_{k=-\infty}^{\infty}\ p(k,n,j;\mu)\ b^{\dag}_{k-n}\
b_k \ , \qquad j\ge 0\ ,
\label{vi}
\end{eqnarray}
where $\ddagger\ \ddagger $ denotes an ordering of the first-quantized
operators $\ \exp (-i\theta)\ $ and $\ i\partial_{\theta}\ $,
such that ${V^j_n}^{\dag}=V^j_{-n}$.
The coefficients $\ p(k,n,j;\mu)\ $ are $j$-th order polynomials
in $k$ whose specific form depends on the choice of ordering
\cite{ctz3}.
The algebra satisfied by these operators follows from the standard
anticommutation rules of the fermionic operators $F$ and
$F^{\dag}$. It is the $\winf$ algebra \cite{shen}, the
quantum version of the algebra of
{\it area-preserving diffeomorphisms}:
\begin{eqnarray}
\left[\ V^i_n\ ,\ V^j_m\ \right] &=& \left(jn-im\right)\ V^{i+j-1}_{m+n}\ +\
q(i,j,m,n)\ V^{i+j-3}_{m+n}\nonumber\\
 &+&\dots\ + \delta_{n+m,0}\ c\left(n,i,j\right)\ .
\label{walgebra}
\end{eqnarray}
The first term in the r.h.s. of (\ref{walgebra}) is the classical
term. It accounts for the {\it local conservation of particle number},
{\it i.e.},
the area under the curve of the density as a function of the spatial
coordinate (which explains the name assigned to (\ref{walgebra})).
The second and higher operator terms arise at the quantum
level because the $V^i_n$ are polynomials in $\ \partial_{\theta}\ $.
Finally, the $c$-number terms $\ c(n,i,j)\ $ represent the relativistic
quantum anomaly, which follows from the renormalization
of the charges $\ V^i_0\ $.
Since we shall measure charges with respect to the ground
state (\ref{vac}),we adopt the standard relativistic normal ordering
procedure of writing all annihilators to the right of creators.

The simplest cases of eq.(\ref{walgebra}) are
\begin{eqnarray}
\left[\ V^0_n,V^0_m\ \right] &=& c\ n\ \delta_{n+m,0}\ ,\nonumber\\
\left[\ V^1_n,V^0_m\ \right] &=& -m\ V^0_{n+m}\ ,\nonumber\\
\left[\ V^1_n,V^1_m\ \right] &=& (n-m)\ V^1_{n+m}\ +\
        {c\over 12}\left(n^3-n\right)\delta_{n+m,0}\ ,
\label{kacmoody}
\end{eqnarray}
with $c=1$.
Eqs.(\ref{kacmoody}) show that $\ V^0_n\ $ and $\ V^1_n\ $ are
the oscillator
(abelian Kac-Moody) and conformal (Virasoro) modes, respectively.
The index $(i+1)$ is the conformal spin of the $V^i_n$ currents and
$n$ is the moding.

The fermionic character of the ground-state (\ref{vac}) when
$N$ is large can now be specified by the infinite set of conditions
\begin{equation}
V^i_n\ \vert\ \Omega\ \rangle\ =\ 0,\qquad \forall\ \
n> 0\ ,i\ge 0\ ,
\label{hwc}
\end{equation}
where $\vert\ \Omega\ \rangle \equiv \vert\ \Omega, N\ \rangle\ $
and $N \gg 1$.
In mathematical terms \cite{bpz}, this equation states that the
ground state is a
{\it c=1 highest-weight state} of the $\ W_{1+\infty}\ $ algebra with
weights $\ V^i_0\vert\ \Omega\ \rangle\ =\ 0\ ,\ \forall i\ge 0\ $.
Moreover, all neutral excitations generated by polynomials of $\ V^i_n\ (n<0)$
applied to $\vert\ \Omega\ \rangle$, make up a
{\it unitary irreducible highest-weight representation of} $W_{1+\infty}$
\cite{kac1}\cite{ctz3}. In a fermionic language, these correspond
to neutral particle-hole excitations on top of $\vert\ \Omega\ \rangle $.
Conditions (\ref{hwc}) are analog to those defining the Laughlin
{\it incompressible quantum fluids} that describe the plateaux
in the quantum Hall effect \cite{prange}\cite{laugh}. Indeed,
this is a consequence of the uniform density  (\ref{gsd}) that
characterizes the ground state $\vert\ \Omega\ \rangle $.
There exist also other types of excitations, corresponding to classical
{\it solitons} \cite{soli}. These are the analogs
of the Laughlin's quasi-holes and quasi-particles in the quantum Hall
effect \cite{laugh}. They carry a non-vanishing
charge $Q$ associated to the {\it local} $U(1)$ particle number conservation
at the Fermi surface \cite{hald1}.
A highest-weight state $|\ Q\ \rangle$ satisfies:
\barr
V^i_n\ |\ Q\  \rangle\ & = &\ 0\ , \qquad \forall\ n>0\ ,\ i \ge 0\ ,
\nonumber\\
V^i_0\ |\ Q\  \rangle\ & = &\ m_i(Q)\ |\ Q\  \rangle\ .
\label{hwsq}
\earr
The weight polynomials $m_i(Q)$ are known \cite{kac1}. For our purposes,
it suffices to quote that $m_0(Q)=Q$ and $m_1(Q)= Q^2/2$. The
ground state $\vert\ \Omega\ \rangle$ is a trivial example of
a highest-weight state with $Q= 0$.
In our algebraic formulation, it can be shown that these charged
excitations, together with their towers of neutral excitations,
correspond to further irreducible highest-weight
representations \cite{ctz3}. All these highest-weight representations
closed under the {\it fusion rules}, {\it i.e.}, the rules for
composition of excitations, define a {\it $\winf$ conformal field theory}.
In physical terms, this is the effective theory of low-energy density
fluctuations about a state with uniform density.

For a highest-weight state $|\ Q\ \rangle$, a (bosonic) basis of neutral
excitations is given by the $V^0_n$ only \cite{kac1}. These states
are of the form:
\beq
|\ k ,\{n_1,n_2,\dots,n_s\}\ \rangle = V^0_{-n_1} V^0_{-n_2} \dots
V^0_{-n_s}|\ Q\ \rangle\ ,\qquad n_1 \ge n_2 \ge \dots \ge n_s > 0\ ,
\label{neut}
\eeq
where $k=Q^2/2 + \sum_{i=1}^s n_i$ is the total momentum (in units
of $2\pi / L$) and $n_i$ are (positive) integers.

In order to finish with the construction of the effective field
theory of (\ref{ham}) and compute the spectrum of its excitations,
we repeat the same steps that brought us to (\ref{efop}) with the
hamiltonian.
Consider the second-quantized free hamiltonian ${\cal H}_0$
(${\cal H}_{CS} =
{\cal H}_0 + {\cal H}_I$)
\beq
{\cal H}_0\ =\ \int_{0}^{L} dx\ {\Psi}^{\dag}(x,0)\ h_0\ \Psi (x,0)\
=\ \sum_{k=-\infty}^{\infty}\ \left( \frac{2\pi}{L} \right)^2
k^2\ a^{\dag}_k a_k\ ,
\label{freeh}
\eeq
with $h_0= -{\partial}^{2}_{x}$.
Following the procedure described above, after subtraction of
the total ground state energy $E_0 = \pi^2 N(N^2-1)/3L^2$
by normal ordering, we find that ${\cal H}_0 = {\cal H}^{(+)}_0
+ {\cal H}^{(-)}_0$ with
\beq
{\cal H}^{(\pm)}_0\ =\ (2\pi \rho_0)^2 \left[\ \frac{1}{4}\ V^0_0\ +\
\frac{1}{N}\ V^1_0\ +\ \frac{1}{N^2}\ \left(\ V^2_0\ -\ \frac{1}{12}\ V^0_0\
\right)\ \right]\ ,
\label{effham}
\eeq
where we have again omitted the $(\pm)$ affix in the $V^i_0$'s
(but it should be clear that both sets of operators are identical
in terms of each Fock basis).
The explicit form of the $\winf$ operators here is \cite{ctz4}:
\barr
V^0_n\ & = &\ \sum_{r=-\infty}^{\infty}\ :b^{\dag}_{r-n} b_{r}:\ ,
\nonumber \\
V^1_n\ & = &\ \sum_{r=-\infty}^{\infty}\
\left( r- {n+1 \over 2}\right)\ :b^{\dag}_{r-n} b_r:\ , \nonumber \\
V^2_n\ & = &\ \sum_{r=-\infty}^{\infty}\
\left( r^2- (n+1)r +{(n+1)(n+2)\over 6} \right)\ :b^{\dag}_{r-n} b_r:\ .
\label{vvv}
\earr
The spectrum of the free hamiltonian follows from (\ref{effham})
and the representation theory of $\winf$ \cite{kac1},
and will be discussed later.

We can also treat the interaction hamiltonian in (\ref{ham})
in a similar way. Consider the second-quantized hamiltonian
\beq
{\cal H}_I\ =\ \int_{0}^{L} dx\ \int_{0}^{L} dy\ {\Psi}^{\dag}(x,0)
{\Psi}^{\dag}(y,0)\
\frac{{\pi}^2}{2 L^2} \frac{g}{\sin^2 ( \pi(x-y)/L)}\
\Psi (y,0) \Psi (x,0)\ .
\label{inth}
\eeq
In Fock space we have
\barr
{\cal H}_I & = &
\sum_{n=-\infty}^{\infty} \sum_{m=-\infty}^{\infty}
\sum_{k=-\infty}^{\infty} {\tilde M}(n,m;k)\
a^{\dag}_{m+k} a^{\dag}_{n-k} a_n a_m\  ,\nonumber \\
{\tilde M}(n,m;k) & \equiv & \frac{1}{2} \left[\ M(n,m;k)\ -\
M(n,m;n-m-k)\ \right]\ , \nonumber \\
M(n,m;k) & \equiv &  g \frac{{\pi}^2}{2 L^2}\
\int_{0}^{L} dx\ \int_{0}^{L} dy\ {\phi}^{*}_{n-k}(x)\
{\phi}_n (x)\ \frac{1}{\sin^2 ( \pi(x-y)/L)}\ \cdot
\nonumber\\
& & \qquad\qquad\qquad{\phi}^{*}_{m+k} (y)\ \phi_m (y)\ .
\label{intfo}
\earr
The formal expression of the integral in (\ref{intfo}) is
ultra-violet divergent, and needs to be regularized.
Explicit computation results in
\beq
M(n,m;k)\ \equiv\ M(|k|)\ =\ g\ \frac{\pi}{4 L^2}
\int_{\epsilon}^{2\pi-\epsilon} d\theta\
\frac{\cos(k\theta)}{\sin^2({\theta/2})}\ ,
\label{mk}
\eeq
where $\epsilon\to 0^+$ is the cut-off.
After regularization, the divergence cancels out in (\ref{intfo})
due to the antisymmetry of the matrix element. The finite
part yields $M(|k|) = - g\ (\pi/L)^2 |k|$.
The hamiltonian\footnote{Note that in rewriting (\ref{intfo})
using the operators $V^0_k$, a subtraction of a term proportional to
the particle number operator is needed \cite{ctz4}. }
in the thermodynamic regime becomes ${\cal H}_I = {\cal H}^{(+)}_I +
{\cal H}^{(-)}_I$, with
\beq
{\cal H}^{(\pm)}_I\ =\ -\ g\ (2\pi \rho_0)^2\ \frac{1}{2 N^2}\
\sum_{k=1}^{\infty}\ k\ V^0_{-k}\ V^0_k\ ,
\label{hiw}
\eeq
where we have omitted again the affix in the operators $V^0_k$, and
the mixed chirality terms in ${\cal H}_I$
have been neglected due to the independence between left and right
chirality sectors assumed in the Hilbert space
of the effective theory.
The {\it effective field theory} of one chiral
sector (say $(+)$) of (\ref{ham}) is finally defined by the
hamiltonian
${\cal H}^{(+)}= {\cal H}^{(+)}_0 + {\cal H}^{(+)}_I$,
with ${\cal H}^{(+)}_0$ given by (\ref{effham}) and
${\cal H}^{(+)}_I$ by (\ref{hiw}).

We are now in a favorable position to discuss the spectrum
of ${\cal H}^{(+)}$. First note that, to order
$1/N$, only the {\it free} hamiltonian (\ref{effham})
contributes. In the basis (\ref{neut}), the finite-size (``gapless'')
spectrum $\Delta {\cal E}_k= {\cal E}_k - (\pi \rho_0)^2 V^0_0$
of (\ref{effham}) is readily obtained:
\barr
\Delta{\cal E}_k\ & = &\ (2\pi \rho_0)^2\ \frac{k}{N}\ +\
O\left(\frac{1}{N^2}\right), \nonumber\\
k\ & = &\ \frac{1}{2}\ Q^2\  +\  \sum_{i=1}^s\ n_i\ ,\qquad n_1 \ge n_2
\ge \dots \ge n_s > 0\ .
\label{specf}
\earr
This leading, conformal part of the spectrum has been already
discussed in \cite{kaya}. The subleading terms of (\ref{effham})
are not diagonal in the basis (\ref{neut}), as opposed to the
interaction terms (\ref{hiw}). However, it is possible,
although tedious, to compute the $1/N^2$ corrections to the
conformal spectrum (\ref{specf}), using (\ref{effham}),(\ref{hiw})
and (\ref{vvv}) (see \cite{ctz4} for details).

Having formulated the free theory in terms of the $\winf$
generators allows us to incorporate the non-perturbative
effects of the interaction by virtue of Luttinger's theorem
\cite{hald1}, which in the $\winf$ language simply amounts
to changing the representation of the symmetry algebra.
To see this, we first
discuss the possible values of $Q$, which could be any real
number according to the representation theory of $\winf$ \cite{kac1},
but it should be fixed by imposing some physical conditions.
In the free theory, $Q$ counts the number of charges
(``soliton number'') added to or removed from the neutral
ground state, and should be, therefore, an integer.
Intuitively, these ``solitons'' correspond to lumps of minimal
size given by the value of the density, appropriately normalized
to $1$.
In a system with a fixed number of particles, as we are considering
throughout this paper, soliton-antisoliton
pairs can be created by coupling the system to an external source
of momentum, which adiabatically adds a ``unit of momentum''
$\Delta p = 2\pi/L$ to every single-particle state. This is in
total analogy with the Laughlin gedanken experiment \cite{laugh},
in which a unit of quantum flux is added to a system of
particles in the first Landau level to create a quasi-particle
or quasi-hole.

Next we note that the Kac-Moody algebra in (\ref{kacmoody})
allows for a rescaling of the operators $V^0_n$
by a real parameter $\xi$, ${\tilde V}^0_n\ =\ {\xi}^{-1}\ V^0_n$,
without modifying the rest
of the $V^i_n$ with $i \ge 1$. This produces a rescaling of the
charge ${\tilde Q}\ =\ {\xi}^{-1}\ Q$.
Moreover, the energy of the state $|\ {\tilde Q}\ \rangle$ with
$n_i=0$ with respect to the (neutral)
ground state $|\  Q=0\ \rangle$ is $\Delta{\tilde {\cal E}}_k=
(2\pi\rho_0/ \xi)^2 Q^2/2N= \Delta{\cal E}_k/{\xi}^2 $, with the last
term given by (\ref{specf}).
It follows that the soliton described by $|\ {\tilde Q}\ \rangle$ can
be reinterpreted as the soliton corresponding to $|\  Q\ \rangle$ if the
density of the system is rescaled as $\rho=\rho_0/\xi$. In intuitive
terms, the normalization of the density is now $1/\xi$ rather than
$1$ and the lumps have minimal height also given by $1/\xi$.
For the class of $(1+1)$-dimensional models in which the ground-state
density scales when varying the coupling constant $g$
(at least in some spatial region), {\it i. e.}, $\rho(g) \xi(g)=
\rho(0) \xi(0)$, we can view the above scaling as the effect of the
{\it interaction}, such that $\xi \equiv \xi (g)$ and normalized as $\xi(0)=1$.
Indeed, theories belonging to this class are known as {\it Luttinger
liquids}, and the property that defines them is known as the ``Luttinger
theorem'' \cite{hald1}. For these systems, the above rescaling can be
considered as the dominant effect of the interaction.
Moreover, on top of each highest-weight state with $Q=\xi(g)m,\ m\in{\bf Z}$,
we always have a tower of neutral excitations as in (\ref{neut}).
These properties fix completely the finite-size spectrum to be
\barr
\Delta {\cal E}_k\ & = &\ \left(\frac{2\pi \rho_0}{\xi(g)} \right)^2\
\frac{k}{N}\ +\ O\left(\frac{1}{N^2}\right)\ , \nonumber \\
k\ & = &\ \frac{1}{2} \left[ m \xi (g) \right]^2\ +\ \sum_{i=1}^s\ n_i\ ,
\quad n_1 \ge n_2 \ge \dots \ge n_s > 0\ ,
\label{confd}
\earr
where $m,n_1,\dots,n_s$ are integers.
The Calogero-Sutherland model belongs to the Luttinger
universality class, with $\xi(g(\lambda))= 1/\sqrt{\lambda}$  \cite{sut}.
This spectrum agrees with the results of \cite{kaya}, and we have also
indicated the structure of the higher order (non-conformal) corrections
to it.
The above result applies assuming that the interaction is turned
on at constant $L$, and during the adiabatic process
an external source or drain adjusts the number of particles
as $N'=N/\sqrt{\lambda}$ to produce the correct value of the density.

Two comments for the specific case of the Calogero-Sutherland model
are in order. The first is that one could realize the
$\winf$ $c=1$ conformal field theory considered here by means of
a chiral boson (see,{\it e.g.}, ref. \cite{fjac}\cite{cdtz}).
In this language, $\sqrt{\lambda}$ is the compactification radius
of the chiral boson that describes the system of interacting fermions
\cite{kaya}.
The second comment is that one could also relate our results
to those that make use of the Bethe ansatz \cite{sut}\cite{ha}.
One can interpret the result (\ref{confd}) as arising from two
contributions: there is a ``collective'' contribution to $k$,
$k_c =m^2/(2\lambda)$ due to solitonic deformations of the density,
and a ``particle-hole'' contribution $k_{ph} = \sum_{i=1}^s n_i$.
In the first-quantized language of \cite{ha},
the former corresponds to the ``total momentum'' of the ground
(or collective) state, whereas the latter
corresponds to the contribution of neutral excitations generated
by Jack polynomials.

A similar analysis can be performed if the
fermions are located on the real line, rather than on a circle.
In this case, the hamiltonian is \cite{cal}\cite{sut}
\beq
{\hat h}_{CS}=\sum_{j=1}^N\ \left[\left( \frac{1}{i}
\frac{\partial}{\partial x_j}\ \right)^2\
+\ \omega^2\ x^2_j\ \right]\ +\
g\ \sum_{j<k}\ \frac{1}{(x_j -x_k)^2}\ .
\label{hamc}
\eeq
Here $\omega$ is the strength of the confining harmonic potential,
and $g$ the coupling constant.
We repeat the construction of the effective theory done for (\ref{ham}).
Consider first
the free case ($g=0$). The single-particle wave functions and spectrum
are given by
\barr
{\hat\varphi}_n(x)\ & = &\ {\cal N}_n\ H_n(\sqrt{\omega} x)\
\exp\left(- \frac{\omega x^2}{2}\right)\ ,
\quad {\cal N}_n = \left( \frac{\omega}{\pi} \right)^{1/4}
\frac{1}{2^{n/2} \sqrt{n!}}\ ,\label{showf} \\
{\hat\epsilon}_n\ & = &\ 2\omega\ \left(\ n\ +\ \frac{1}{2} \right)\ ,
\qquad n=0,1,2, \dots\ ,
\label{hermi}
\earr
where $H_n(x)$ are the Hermite polynomials. For convenience, we also
define ${\hat\varphi}_n(x,t) \equiv {\hat\varphi}_n(x) \exp(-i
{\hat\epsilon}_n t)$. Next, we introduce a second quantized field
\beq
{\hat \Psi} (x,t)\ =\ \sum_{n=0}^{\infty} a_n\ {\hat\varphi}_n (x,t)\ ,
\qquad \{\ a_k , a^{\dag}_l\ \}\ =\ \delta_{k,l}\ .
\label{fielc}
\eeq
The ground state is
\beq
|\ \Omega\ \rangle\ =\ a^{\dag}_{N-1} a^{\dag}_{N-2} \dots
a^{\dag}_1 a^{\dag}_0\ |\ 0\ \rangle\ .
\label{gscal}
\eeq
The ground state density $\rho(x)$ (defined as in (\ref{gsd}))
in the thermodynamic limit satisfies the {\it semicircle law}
(see, {\it e.g.},\cite{sut}):
\beq
\rho(x)\ =\ \frac{\omega}{\pi}\ \sqrt{x^2_0\ -\ x^2\ }\ ,\qquad
x^2_0\ =\ \frac{2N}{\omega}\ ,
\label{semi}
\eeq
for values $x^2 \le x^2_0$ and zero otherwise. The semiclassical
location of the spatial boundary of the ground state is given by
$\pm x_0$, and the (semiclassical) value of the Fermi momentum is
$p_F=\sqrt{2N\omega}$.
The thermodynamic limit of $N$ large and $\omega$ small,
is taken such that $p_F$ and the value of the density
at the origin $\rho(0)=\sqrt{2N\omega}\ / \pi$ are finite quantities.
We then search for the smallest energy excitations, located
in regions in which the density (\ref{semi}) is approximately
constant, {\it i.e.}, around the origin. Define a coordinate
$q \equiv x/x_0$ and expand $\rho(x)$ around $x=0$.
We can turn this small $q$ expansion into a consistent $1/N$
expansion by considering values of $q$ such that $q^2 \simeq
O(1/N^2)$ only. For excitations close to the Fermi surface,
the kinetic term in (\ref{hamc}) can be estimated semiclassically
to be of the order of $p^2_F \simeq O(1)$. However, the
confining potential term in (\ref{hamc}) is of
the order of $p^2_F q^2 \simeq O(1/N^2)$. Therefore, in this
spatial region the system behaves like in the compact
case of (\ref{ham}), and one could repeat the analysis previously
done for that case.

In order to define the theory of small fluctuations about the Fermi
surface, consider again the ``shifted'' Fock operators
$b_r = a_{N-1+r}$, with $|r| \ll N$. We consider values of $N\gg 1$
in (\ref{fielc}), with $q^2 \simeq O(1/N^2)$ only.
The asymptotic form of the wave functions (\ref{showf}) for
$n \gg 1$ under this assumption is given by \cite{bate}
\beq
{\hat\varphi}_n(x)\ =\ \frac{1}{\sqrt{\pi}}\ \left( \frac{2\omega}
{n} \right)^{1/4}\ \left[\ \cos \left( \sqrt{\omega(2n+1)}\ x\ -\
n\frac{\pi}{2}\ \right)\ +\ O\left( \frac{1}{N^2} \right)\ \right]\ ,
\label{asywf}
\eeq
where the last term is proportional to $q^2$.
After rescaling $x$ and $t$ as $x=q x_0$ and $t=\tau x_0/2$,
and following the steps that brought us to (\ref{efop}), we find
\barr
{\hat \Psi} (x,t)\ & = &\ \frac{1}{2\sqrt{\pi}} \left(\ \frac{2\omega}
{N} \right)^{1/4}\ \left[\ {\rm e}^{i\beta(+)}
{\hat F}^{(+)} (q-p_F\tau)\ +\ {\rm e}^{-i\beta(-)}\
{\hat F}^{(-)} (q+p_F\tau)\  \right]\ , \nonumber \\
{\hat F}^{(\pm)}(\theta)\ & = &\ \sum_{r=-\infty}^{\infty} {\tilde b}_r\
{\rm e}^{\pm i(r-1/2)\theta}\ , \qquad
\beta (\pm)\ \equiv\ 2N(q \mp  p_F\tau/2) \mp (N-1)\frac{\pi}{2}\ ,
\label{efopca}\earr
where ${\tilde b}_r \equiv \exp(-ir\pi/2)\ b_r $ . This result
is in agreement with (\ref{efop}) as expected from the previous
intuitive discussion.
Similarly, the second quantized free hamiltonian ${\hat{\cal{H}}}_0$
(${\hat{\cal{H}}}_{CS} = {\hat{\cal{H}}}_0 + {\hat{\cal{H}}}_I$) becomes
\beq
{\hat{\cal{H}}}_0\ =\ \int_{-\infty}^{\infty} dx\
{\hat\Psi}^{\dag}(x,0)\ {\hat h}_0\ {\hat\Psi} (x,0)\
=\ \sum_{n=0}^{\infty}\ 2\omega\ \left(\ n\ +\ \frac{1}{2}\ \right)\
a^{\dag}_n a_n\ ,
\label{fcah}
\eeq
with ${\hat h}_0= -{\partial}^{2}_{x}+ {\omega}^2 x^2$.
After subtraction of the total ground state energy
${\hat E}_0 = \omega N^2$ by normal ordering, we find
\beq
{\hat{\cal{H}}}_0\ =\ 2\omega N
\left( V^0_0\ +\ \frac{1}{N} V^1_0\ \right)\ .
\label{effhca}
\eeq
Had we considered a chiral splitting as in the compact case of (\ref{ham}),
we would have found ${\hat{\cal H}}_0 = {\hat{\cal H}}^{(+)}_0 +
{\hat{\cal H}}^{(-)}_0$ with ${\hat{\cal H}}^{(+)}_0={\hat{\cal H}}^{(-)}_0$
as in (\ref{effham}). However,
the correct normalization of the chiral part of (\ref{asywf}) gives
for ${\hat{\cal H}}^{(\pm)}_0$ exactly the same result of (\ref{effhca}).
In the thermodynamic limit, both hamiltonians (\ref{effham}) and
(\ref{effhca}) belong to the same universality class, as expected.

To include the effects of the interaction, consider the
second quantized hamiltonian
\beq
{\hat{\cal{H}}}_I\ =\ \frac{1}{2}\ \int_{-\infty}^{\infty} dx\
\int_{-\infty}^{\infty} dy\
{\hat\Psi}^{\dag}(x,0){\hat\Psi}^{\dag}(y,0)\
\frac{g}{(x -y)^2}\ {\hat\Psi}(y,0) {\hat\Psi}(x,0)\ .
\label{inca}
\eeq
In Fock space we have an expression analog to (\ref{intfo}),
but with two-body matrix elements of the form
\beq
M(n,m;k) \equiv \frac{1}{2}\ \int_{-\infty}^{\infty} dx\
\int_{-\infty}^{\infty} dy\
{\hat\varphi}^{*}_{n-k}(x)\ {\hat\varphi}_n (x)\
\frac{g}{(x -y)^2}\ {\hat\varphi}^{*}_{m+k}(y)\
{\hat\varphi}_m (y)\ ,
\label{incal}
\eeq
and $n \ge 0$, $m \ge 0$ and $-m\ \le\ k\ \le\ n$ .
In practice, the only values of $n,m$ we are interested in
are of $O(N)$.
The wave functions are given by (\ref{asywf}), but we shall
be concerned with one chirality (say $(+)$) only:
\beq
{\hat\varphi}^{(+)}_{N+r}(x)\ =\ \sqrt{ \frac{2}{\pi} }\
\left(\ \frac{2\omega}{N} \right)^{1/4}\ \exp \left[\
i\sqrt{2\omega N} x + i\sqrt{ \frac{\omega}{2N} } \left( r+ \frac{1}{2}
\right) x -i(N+r)\frac{\pi}{2}\ \right]\ .
\label{appwf}
\eeq
After inserting this effective (correctly normalized) wave function in
(\ref{incal}), and
consistently restricting the domain of integration to $-L_{eff} \le
x,y \le L_{eff}$, with $L_{eff} \equiv \pi x_0$, we follow the
procedure that brought us to (\ref{hiw}). In the thermodynamic limit,
the explicit result for the interaction hamiltonian is
\barr
{\cal H}_I\ & = &\ -\ g\ \frac{8}{{\pi}^2}\ 2\omega N\ \frac{1}{N^2}\
\sum_{k=1}^{\infty} f(k)\ V^0_{-k}\ V^0_k\ ,\nonumber\\
f(k)\ & = &\
\int_0^{\pi}\ dx\ \left(\pi-x\right)\ \frac{(1 - \cos (2k x))}{x^2}\
\simeq\  {\pi}^2\ k\ ,\ \sqrt{N} \gg k \gg 1\ .
\label{mkp}
\earr
Once more, this term is of
order $1/N^2$ in the thermodynamic regime and, therefore, negligible
to first order  in $1/N$.

Having formulated the theory in terms of the $\winf$
generators allows us to invoke Luttinger's theorem again.
As a consequence of it, and eqs. (\ref{effhca}) and (\ref{mkp}),
the finite-size spectrum and conformal dimensions
are given again by (\ref{confd}) with $\xi(g)= 1$ \cite{sut},
replacing $(2\pi \rho_0)^2$ by $2\omega N$.
In deriving this result, we have used the same definitions
and normalizations as in the compact case (in particular, the
value of $\rho(0)$ remains fixed even in the presence of
interaction to account for the change in the total
particle number).
Therefore, we have explicitly shown that the $\winf$
bosonization scheme gives the same universal results
in both the cases of the circle (eq. (\ref{ham})) and
the infinite line (eq. (\ref{hamc})).

Before concluding, we would also like to explain how the
$\winf$ approach naturally leads to a {\it classification} of
universality classes of non-relativistic (chiral) fermionic
systems by focusing on the dynamics of the Fermi surface.
In a general chiral system, the $\winf$ structure will be
automatically incorporated in the Fermi sea and
a generic (low-energy) effective hamiltonian will have an
expansion of the form
\beq
{\cal H}\ =\ \sum_{i}\ \alpha_i\ V^i_0\ +\
\sum_{i,j,n}\ \beta_{ij}(n)\ V^{i}_{-n}\ V^{j}_n\ ,
\label{hexp}
\eeq
where $\alpha_i$ and $\beta_{ij}(n)$ are the coefficients that
contain all the relevant information about the effective theory,
the former arising from one-body terms and the latter from
two-body interactions in the microscopic theory \cite{ctz4}.
A systematic study of the possible universality classes
compatible with (\ref{hexp}) can be done using the powerful
and general mathematical results of the representation theory of
$\winf$ \cite{kac1}. For example, we expect that those $\winf$
conformal field theories with $c \ge 2$ ($c$ is constrained to be
a natural number by unitarity \cite{kac1}) will exhibit a non-trivial
Fermi surface structure reminiscent of the hierarchy in the
quantum Hall effect \cite{ctz3}.

We would like to thank M. Frau and S. Sciuto for many useful
discussions and for comments on the manuscript. G. R. Z. would
like to acknowledge early discussions on related subjects and
the many insights on the representation theory of $\winf$ that
A. Cappelli and C. A. Trugenberger generously shared with him.
%
\def\NP{{\it Nucl. Phys.\ }}
\def\PRL{{\it Phys. Rev. Lett.\ }}
\def\PL{{\it Phys. Lett.\ }}
\def\PR{{\it Phys. Rev.\ }}
\def\IJMP{{\it Int. J. Mod. Phys.\ }}
\def\MPL{{\it Mod. Phys. Lett.\ }}


\begin{thebibliography}{99}


\bibitem{cal}   F. Calogero, {\it J. Math. Phys.} {\bf 10} (1969) 2191,
                {\it ibid.} {\bf 10} (1969) 2197, {\it ibid.} {\bf 12}
                (1971) 419.
\bibitem{sut}   B. Sutherland, {\it J. Math. Phys.} {\bf 12} (1971) 246,
                {\it ibid} {\bf 12} (1971) 251,
                {\it Phys. Rev.} {\bf A 4} (1971) 2019, {\it ibid.} {\bf A 5}
                (1972) 1372.
\bibitem{hald2} F. D. M. Haldane, \PRL {\bf 67} (1991) 937.
\bibitem{csw}   A. P. Polychronakos, \PRL {\bf 69} (1992) 703;
                K. Hikami and M. Wadati, {\it J. Phys. Soc. Jpn.}
                {\bf 61} (1992) 3425, {\it ibid.} {\bf 62} (1993) 3035,
                {\it ibid.} {\bf 62} (1993) 4203;
                D. V. Khveschenko, preprints cond-mat/9401012
                and cond-mat/9404094;
                S. Iso, preprint hep-th/9411051;
                S. Iso and S.J. Rey, preprint hep-th/9406192;
                E. Bergshoeff and M. Vasiliev, preprint hep-th/9411093;
                H. Awata, Y. Matsuo, S. Odake and J. Shiraishi,
                \PL {\bf B 347} (1995) 49, and preprints hep-th/9503028 and
                hep-th/9503043;
                D. Bernard, K. Hikami and M. Wadati, preprint hep-th/9412194.
\bibitem{chain} F. D. M. Haldane, \PRL {\bf 60} (1988) 635; B. S. Shastry,
                \PRL {\bf 60} (1988) 639.
\bibitem{qcd}   M. R. Douglas, preprint hep-th/9303159;
                M. Caselle, A. D'Adda, L. Magnea and S. Panzeri, \NP
                {\bf B 416} (1994) 751;
                J. A. Minahan and A. P. Polychronakos, \PL {\bf B 312}
                (1993) 155;
                A. Gorskii and N. Nekrasov, \NP {\bf B 436} (1995) 582;
                J. A. Minahan and A. P. Polychronakos, \PL {\bf B 326}
                (1994) 288.
\bibitem{anyon} T. H. Hansson, J. M. Leinaas and J. Myrheim, \NP
                {\bf B 384} (1992) 559; L. Brink, T. H. Hansson,
                S. Konstein and M. A. Vasiliev, \NP {\bf B 401} (1993)
                591; S. B. Isakov, \IJMP {\bf A 9} (1994) 2563.
\bibitem{quaw}  M. Caselle, preprint DFTT 52/94 cond-mat/9410097.
\bibitem{csqhe} M. Stone and M. Fisher, \IJMP {\bf B 8} (1994) 2539;
                S. Iso, \MPL {\bf A 9} (1994) 2123; H. Azuma and S. Iso,
                \PL {\bf B 331} (1994) 107.
\bibitem{jevic} I. Andri\'c, A. Jevicki and H. Levine, \NP {\bf B 215} (1983)
                307; I. Andri\'c and V. Bardek, {\it J. Phys.} {\bf A 21}
                (1988) 2847; J. Avan and A. Jevicki, \PL {\bf B 266} (1991)
                35, and {\it Comm. Math. Phys.} {\bf 150} (1992) 149;
                A. Jevicki, \NP {\bf B 376} (1992) 75.
\bibitem{polch} See, for example: J. Polchinski, {\it ``Effective field
                theory and the Fermi surface''}, Lectures presented at
                TASI-92, Boulder (U.S.A.), June 1992, preprint hep-th/9210046.
\bibitem{cdtz}  A. Cappelli, G. V. Dunne, C. A. Trugenberger and G. R.
		Zemba, \NP {\bf 398 B} (1993) 531.
\bibitem{ctz4}  A. Cappelli, C. A. Trugenberger and G. R. Zemba,
		preprint cond-mat/9407095.
\bibitem{hald1} F. D. M. Haldane, {\it ``Luttinger's Theorem and Bosonization
                of the Fermi surface''}, lectures given at the
                {\it International School of Physics Enrico Fermi},
                Varenna, Italy (July, 1992), and {\it J. Phys.} {\bf C 14}
                (1981) 2585; for the chiral case, X.-G. Wen, \PR {\bf B 40}
                (1989) 7387.
\bibitem{bpz}   A. A. Belavin, A. M. Polyakov and A. B. Zamolodchikov,
		\NP {\bf B 241} (1984) 333; for a review see:
		P. Ginsparg, {\it Applied Conformal Field Theory},
		in {\it Fields, Strings and Critical Phenomena},
		Les Houches School 1988, E. Brezin and J. Zinn-Justin eds.,
		North-Holland, Amsterdam (1990).
\bibitem{boson} For formulations of bosonization that are closer to
                our ideas, see A. H. Castro Neto and E. Fradkin,
                \PRL {\bf 72} (1994) 1393, \PR {\bf B 49} (1994) 10877;
                A. Houghton and B. Marston, \PR {\bf B 48} (1993) 7790.
\bibitem{kaya}  N. Kawakami and S.-K. Yang, \PRL {\bf 67} (1991) 2493,
                {\it Prog. Theor. Phys. Suppl.} {\bf 107} (1992) 59.
\bibitem{prange} For a review see: R. A. Prange, S. M. Girvin, {\it The Quantum
		Hall Effect}, Springer Verlag, New York (1990).
\bibitem{shen}  I. Bakas, \PL {\bf B 228} (1989) 57;
		C. N. Pope, X. Shen and L. J. Romans, \NP {\bf B 339} (1990);
		for a review see: X. Shen, \IJMP  {\bf 7 A} (1992) 6953.
\bibitem{laugh} R. B. Laughlin, \PRL {\bf 50} (1983) 1395;
		for a review see: R. B. Laughlin, {\it Elementary Theory: the
		Incompressible Quantum Fluid}, in \cite{prange}.
\bibitem{soli}  B. Sutherland and J. Campbell, \PR {\bf B 50} (1994) 888;
                A. P. Polychronakos, preprint hep-th/9411054;
                I. Andri\'c, V. Bardek and L. Jonke, preprint hep-th/9411136.
\bibitem{ha}    Z.N.C.~Ha, \PRL {\bf 73} (1994) 1574; \NP {\bf B 435}
                (1995) 604.
\bibitem{fjac}  R. Floreanini and R. Jackiw, \PRL {\bf 59} (1987) 1873.
\bibitem{ctz3}  A. Cappelli, C. A. Trugenberger and G. R. Zemba,
		\PRL {\bf 72} (1994) 1902; see also preprint hep-th/9502021.
\bibitem{kac1}  V. Kac and A. Radul, {\it Comm. Math. Phys.} {\bf 157}
		(1993) 429; E. Frenkel, V. Kac, A. Radul and W. Wang,
		preprint hep-th/9405121; see also H. Awata, M. Fukuma,
                S. Odake, Y-H. Quano, {\it Lett. Math. Phys} {\bf 31}
                (1994) 289; H. Awata, M. Fukuma, Y. Matsuo and S. Odake,
                \PL {\bf B 332} (1994) 336, preprints hep-th/9405093,
                hep-th/9406111 and hep-th/9408158.
\bibitem{bate}  H. Bateman, {\it Higher Transcendental Functions},
		Vol. II, McGraw-Hill, New York (1953).
\end{thebibliography}
\end{document}